\documentclass[preprint,amsmath,amssymb,aps]{revtex4-2}

\usepackage{graphicx}
\usepackage{dcolumn}
\usepackage{bm}
\usepackage{xcolor}
\usepackage{amsfonts}
\usepackage{graphics}
\usepackage{hyperref}
\usepackage{slashed}
\usepackage{babel}
\makeatletter

\begin{document}

\title{Two-loop superfield effective potential for a $\mathcal{N}=2$, $d=3$ supersymmetric quantum electrodynamics with higher derivatives}

\author{F. S. Gama}
\email{fgama@unifap.br}
\affiliation{Departamento de Ci\^{e}ncias Exatas e Tecnol\'{o}gicas, Universidade Federal do Amap\'{a}, 68903-419, Macap\'{a}, Amap\'{a}, Brazil}

\begin{abstract}
In the $\mathcal{N}=2$, $d=3$ superspace, we consider a higher-derivative generalization of the supersymmetric quantum electrodynamics, where the higher-derivative operator is a polynomial function of the d'Alembertian with arbitrary degree. For this theory, we use the background field quantization in a higher-derivative $R_\xi$ gauge to explicitly calculate the superfield effective potential up to two loops in the K\"{a}hlerian approximation. This superfield effective potential is obtained in a closed form and in terms of elementary functions.
\end{abstract}

\maketitle
\newpage

\section{Introduction}

The higher-derivative (HD) generalization of a given lagrangian is a quite old idea going back to Ostrogradsky's work \cite{OW}. In the 1940s, HD modifications to Maxwell electrodynamics were suggested by Bopp \cite{Bopp} and Podolsky \cite{Podolsky} with the aim to prevent singularities due to point charges. Despite being an old idea, HD theories continue to be investigated to this day for three reasons. First, these theories are unavoidable, in the sense that they arise naturally in different contexts, such as counterterms necessary to ensure the renormalizability of semiclassical gravity theories \cite{Shapiro}, in the small slope expansion of string models \cite{string}, and by integrating out heavy fields in the effective field theory approach \cite{EFT}. Second, HD theories are better behaved than the standard field theories when it comes to classical and quantum divergences. For example, HD gravitational theories of order higher than four have a regular Newtonian potential and no curvature singularities \cite{newtonsing}. Moreover, it is well-known that HD theories of gravity are (super-)renormalizable \cite{QG}. Third, field theories can be regularized by means of the method of higher covariant derivatives \cite{Slavnov}, which is a regularization scheme that preserves gauge invariances as well as supersymmetry \cite{KW}. Recently, this scheme has been employed in the perturbative computation of the anomalous dimension and beta function for supersymmetric quantum electrodynamics (SQED) and minimal supersymmetric standard model \cite{Stepanyantz}.

In this context of HD theories, we will investigate the effective potential (EP). Physically, the EP takes into account the effect of the quantum fluctuations on the classical theory, so that it is a quantum generalization of the classical potential \cite{EP}. The EP is not only important for its own sake, but it is also necessary to study spontaneous symmetry breaking produced by radiative corrections \cite{breaking}, the symmetry restoration at high temperatures \cite{restoration}, and the false vacuum decay \cite{falsevacuum}. Very recently, the gauge dependence of the EP for the standard model of electroweak unification has been investigated in \cite{Nielsen}, the renormalization-group-improved EP has been studied in the Gross-Neveu model \cite{QMF}, and the one-loop EP has been calculated for the scalar-tensor gravity in Ref. \cite{ALN}. 

In supersymmetric field theories, the focus is shifted from the EP to the superfield effective potential (SEP), which is a supersymmetric generalization of the former one \cite{BKY}. The reason for this is that the SEP is defined by means of the superfield formalism in which the supersymmetry invariance is manifest. In view of the significance of the (superfield) effective potential and the excellent properties of HD models discussed above, many studies of the SEP were reported in the literature for different HD extensions of supersymmetric theories, such as chiral superfield models \cite{SHD}, gauge superfield theories \cite{gaugehd,gaugehd2}, and three-dimensional superfield models \cite{GNP,GGNPS}. However, all these studies were limited to the investigation of the one-loop SEP. Recently, we have made progress towards the explicit calculation of the two-loop SEP for HD scalar and chiral superfield theories \cite{GNP2}, which are non-gauge theories. The aim of the present paper is to continue this investigation by computing the two-loop correction to the SEP in a HD version of the $\mathcal{N}=2,d=3$ SQED, which is an abelian gauge theory. It is important to point out that we have already calculated the one-loop SEP of this model in our earlier paper \cite{GGNPS}. Nonetheless, the one-loop SEP was given in terms of integrals over the momenta. Thus, we also aim to improve the result of \cite{GGNPS} by obtaining an explicit expression for the one-loop SEP in this work.

The structure of the paper is as follows. In section II, we introduce the HD version of the $\mathcal{N}=2,d=3$ SQED coupled to chiral matter and develop the background field quantization necessary for performing quantum calculations. In section III, we use the fundamental theorem of algebra together with the partial fraction representation of the propagators to explicitly calculate the SEP up to the two-loop level. In section IV, we give a short summary of the main results obtained and suggest a possible continuation of this study.

\section{$\mathcal{N}=2$, $d=3$ SQED with higher derivatives and background field quantization}

In the $\mathcal{N}=2$, $d=3$ superspace, the SQED describes a gauge superfield $V$ which mediates the interaction between matter superfields $\Phi_+$ and $\Phi_-$  with non-zero $U(1)$ charge \cite{N2SQED}. One possible way to generalize this theory is to include higher derivatives only in the gauge sector
\begin{equation}
	\label{model}
	S_{HD}=\int d^7z\left(-\frac{1}{8g^2}Gf(\Box)G+\bar{\Phi}_+e^V\Phi_++\bar{\Phi}_- e^{-V}\Phi_-\right).
\end{equation}
Here, $G=\bar{D}^\alpha D_\alpha V$ is the gauge invariant field strength. The dimensionless scalar operator $f(\Box)$ is assumed to be a polynomial function with arbitrary degree. In principle, we could also introduce HD operators into the matter sector and add to (\ref{model}) superpotential terms involving only chiral superfields. At the one-loop level, the calculations would be feasible and they would probably go in a very similar way as we have done in \cite{gaugehd2}. On the other hand, the two-loop calculations would be quite demanding due to the more involved structure of the propagators and the large number of vertices. We plan to tackle this problem in future work. In the present contribution, for the sake of simplicity, the HD generalization of the SQED that we will study is the one defined in (\ref{model}).

All quantum calculations in this work will be performed using the background field approach \cite{qb,susyqb}. Following this formalism, we make the quantum/background splitting of the linear form:
\begin{equation}
	\label{split}
	V\rightarrow V+v; \ \ \ \Phi_\pm\rightarrow\Phi_\pm+\phi_\pm; \ \ \ \bar{\Phi}_\pm\rightarrow\bar{\Phi}_\pm+\bar{\phi}_\pm.
\end{equation}
We know by definition that the SEP depends only on chiral and antichiral superfields and their derivatives, so we must set the background gauge superfield to zero. Additionally, we will study the effective action in the so-called K\"{a}hlerian approximation in which the derivatives of the background chiral and antichiral superfields are neglected \cite{BKY}. Thus, we impose on the background superfields the constraints:
\begin{equation}
	\label{constr}
	V=0; \ \ \ D_\alpha\Phi_\pm=0; \ \ \ \bar{D}_{\dot{\alpha}}\bar{\Phi}_\pm=0; \ \ \ \partial_{\alpha\dot{\alpha}}\Phi_\pm=0; \ \ \ \partial_{\alpha\dot{\alpha}}\bar{\Phi}_\pm=0.
\end{equation}
Therefore, inserting Eqs. (\ref{split}) into (\ref{model}), we obtain
\begin{align}
	\label{splited}
	S_{split}&=\int d^7z\bigg[-\frac{1}{8g^2}G f(\Box)G+\left(\bar{\Phi}_++\bar{\phi}_+\right)e^v\left(\Phi_++\phi_+\right)\nonumber \\
	&+\left(\bar{\Phi}_-+\bar{\phi}_-\right)e^{-v}\left(\Phi_-+\phi_-\right)\bigg].
\end{align}
Now, $G=\bar{D}^\alpha D_\alpha v$ and we have two sets of superfields. Consequently, the action is invariant under two types of transformations:
\begin{align}
	\label{background}
	\text{Background:} \ \ &\Phi_{\pm}^\prime=e^{\pm iK}\Phi_{\pm}; \ \ \bar{\Phi}_{\pm}^\prime=e^{\mp iK}\bar{\Phi}_{\pm}; \ \ \phi_{\pm}^\prime=e^{\pm iK}\phi_{\pm}; \ \ \bar{\phi}_{\pm}^\prime=e^{\mp iK}\bar{\phi}_{\pm}; \ \ v^\prime=v;\\
	\label{quantum}
	\text{Quantum:} \ \ &\Phi_{\pm}^\prime=\Phi_{\pm}; \ \ \bar{\Phi}_{\pm}^\prime=\bar{\Phi}_{\pm}; \ \ \phi_{\pm}^\prime=e^{\pm i\Lambda}\left(\Phi_{\pm}+\phi_{\pm}\right)-\Phi_{\pm};\nonumber\\
	&\bar{\phi}_{\pm}^\prime=e^{\mp i\bar{\Lambda}}\left(\bar{\Phi}_{\pm}+\bar{\phi}_{\pm}\right)-\bar{\Phi}_{\pm}; \ \ v^\prime=v+i\left(\bar{\Lambda}-\Lambda\right).
\end{align}
It is worth pointing out that the background transformations are global because the background gauge superfield was set to zero.

To calculate the SEP at the two-loop level, it is sufficient to expand (\ref{splited}) around the background superfields up to the fourth order in the quantum superfields. Therefore, we get
\begin{equation}
	\label{perturbation}
	S_{split}=S^{(0)}+S^{(1)}+S^{(2)}+S^{(3)}+S^{(4)}.
\end{equation}
The first term is just the classical contribution to the effective action, and the second one only
leads to one-particle reducible supergraphs \cite{GGRS}. Therefore, these two terms can be dropped out. On the other hand, the quadratic terms in the quantum superfields are given by
\begin{equation}
	\label{quadratic}
	S^{(2)}=\int d^7z\left\{\frac{1}{8g^2}v\left[f(\Box)\left(-\Box+\{D^2,\bar{D}^2\}\right)+M^2_v\right]v+\bm{\bar{\phi}}\bm{\phi}+v\bm{\bar{\Phi}}\sigma_3\bm{\phi}+v\bm{\bar{\phi}}\sigma_3\bm{\Phi}\right\},
\end{equation}
where $\sigma_3$ is a Pauli matrix and we have introduced the following matrix notation:
\begin{equation}
	\label{matrix_not}
	\bm{\bar{\Phi}}=\left(\begin{array}{cc}
		\bar{\Phi}_+ & \bar{\Phi}_- 
	\end{array}\right); \ \ \ \bm{\Phi}=\left(\begin{array}{c}
		\Phi_+ \\
		\Phi_-
	\end{array}\right); \ \ \ \bm{\bar{\phi}}=\left(\begin{array}{cc}
		\bar{\phi}_+ & \bar{\phi}_- 
	\end{array}\right); \ \ \ \bm{\phi}=\left(\begin{array}{c}
		\phi_+ \\
		\phi_-
	\end{array}\right). 
\end{equation}
Moreover, the mass parameter $M_v^2$ is defined as
\begin{equation}
	\label{mass_gauge}
	M_v^2=4g^2\bm{\bar{\Phi}}\bm{\Phi}.
\end{equation} 
In the standard SQED ($f(\Box)=1$), the parameter $M_v^2$ is the mass acquired by the gauge superfield $v$ due to the non-vanishing background defined by $\bm{\Phi}$. In the HD version of the SQED, the same interpretation still holds although the mass acquired by the gauge superfield is not given by (\ref{mass_gauge}) anymore, it now also depends on the mass scales introduced by means of the operator $f(\Box)$ [see Eq. (\ref{masses})].

The cubic and quartic terms are given by
\begin{align}
	\label{cubic}
	S^{(3)}&=\int d^7z\left[v\bm{\bar{\phi}}\sigma_3\bm{\phi}+\frac{1}{2}v^2(\bm{\bar{\Phi}}\bm{\phi}+\bm{\bar{\phi}}\bm{\Phi})+\frac{1}{3!}v^3\bm{\bar{\Phi}}\sigma_3\bm{\Phi}\right];\\
	\label{quartic}
	S^{(4)}&=\int d^7z\left[\frac{1}{2}v^2\bm{\bar{\phi}}\bm{\phi}+\frac{1}{3!}v^3(\bm{\bar{\Phi}}\sigma_3\bm{\phi}+\bm{\bar{\phi}}\sigma_3\bm{\Phi})+\frac{1}{4!}v^4\bm{\bar{\Phi}}\bm{\Phi}\right].
\end{align}
Since the kinetic operator of the gauge superfield is not invertible (\ref{splited}), we must remove its degeneracy fixing the quantum gauge symmetry (\ref{quantum}), but preserving the background invariance (\ref{background}). This can be achieved by adding to the action the higher-derivative generalization of the usual supersymmetric $R_\xi$ gauge \cite{OW2}
\begin{equation}
	\label{gft}
	S_{GF}=-\frac{1}{4g^2}\int d^7z\bar{F}f(\Box)F,
\end{equation}
where the suitable gauge-fixing function is defined as
\begin{equation}
	\label{gff}
	F=\bar{D}^2\left(v+4g^2\frac{1}{\Box f(\Box)}\bm{\bar{\phi}}\sigma_3\bm{\Phi}\right),
\end{equation}
which was defined in this way with the aim of canceling the unwanted mixing of matter and gauge quantum superfields in (\ref{quadratic}). Substituting (\ref{gff}) into (\ref{gft}), we obtain the following explicit form:
\begin{equation}
	\label{gf_2}
	S_{GF}=-\int d^7z\left[\frac{1}{4g^2}(D^2v)f(\Box)\bar{D}^2v+v\bm{\bar{\phi}}\sigma_3\bm{\Phi}+v\bm{\bar{\Phi}}\sigma_3\bm{\phi}+\bm{\bar{\phi}}\bm{M}^2\frac{1}{\Box f(\Box)}\bm{\phi}\right],
\end{equation}
where we have introduced the mass matrix $\bm{M^2}=4g^2\sigma_3\bm{\Phi}\bm{\bar{\Phi}}\sigma_3$, or
\begin{equation}
	\bm{M^2}=4g^2\left(\begin{array}{cc}
		\left|\Phi_+\right|^2 & -\Phi_+\bar{\Phi}_-\\
		-\Phi_-\bar{\Phi}_+ & \left|\Phi_-\right|^2
	\end{array}\right). 
\end{equation}
The inconvenience of the gauge choice (\ref{gff}) is that the Faddeev-Popov ghosts will interact with
the background superfields, so that they must be included to maintain the consistency of the gauge-fixing procedure. Thus, besides Eq. (\ref{gft}), we also have to add to the model the ghost action
\begin{equation}
	\label{ghost}
	S_{FP}=\left.\left(i\int d^5zc^\prime\delta_\Lambda F+i\int d^5\bar{z}\bar{c}^\prime\delta_\Lambda\bar{F}\right)\right|_{\Lambda\to c, \ \bar{\Lambda}\to \bar{c}}.
\end{equation}
The variation $\delta_\Lambda F$ is obtained from the infinitesimal version of the quantum gauge transformations (\ref{quantum}). Therefore, it is possible to show that
\begin{equation}
	\delta_\Lambda F=i\bar{D}^2\left[\left(1-\frac{M^2_v}{\Box f(\Box)}\right)\bar{\Lambda}-4g^2\frac{1}{\Box f(\Box)}\left(\bar{\Lambda}\bm{\bar{\phi}}\right)\bm{\Phi}\right].
\end{equation}
Substituting this into (\ref{ghost}), we find the following quadratic and cubic contributions
\begin{align}
	\label{ghost_2}
	S_{FP}^{(2)}&=\int d^7z\left[-c^\prime\left(1-\frac{M^2_v}{\Box f(\Box)}\right)\bar{c}+\bar{c}^\prime\left(1-\frac{M^2_v}{\Box f(\Box)}\right)c\right];\\
	\label{ghost_3}
	S_{FP}^{(3)}&  =4g^2\int d^7z\left[c^\prime\frac{1}{\Box f(\Box)}\left(\bar{c}\bm{\bar{\phi}}\right)\bm{\Phi}-\bm{\bar{\Phi}}\bar{c}^\prime\frac{1}{\Box f(\Box)}\left(c\bm{\phi}\right)\right].
\end{align}
Finally, all of the functionals which are quadratic in the quantum superfields (\ref{quadratic}), (\ref{gf_2}), and (\ref{ghost_2}) can be combined into a single expression 
\begin{equation}
	\label{combined}
	S^{(2)}+S_{GF}+S_{FP}^{(2)}=\frac{1}{2}v\cdot\mathcal{H}_v\cdot v+\frac{1}{2}\left(\begin{array}{cc}
		\bm{\phi^T} & \bm{\bar{\phi}}\end{array}\right)\cdot\mathcal{H}_\phi\cdot\left(\begin{array}{c}
		\bm{\phi}\\
		\bm{\bar{\phi}^T}\end{array}\right)+\frac{1}{2}\left(\begin{array}{cccc}
		c & c^\prime & \bar{c} & \bar{c}^\prime\end{array}\right)\cdot\mathcal{H}_{FP}\cdot\left(\begin{array}{c}
		c \\
		c^\prime \\
		\bar{c}\\
		\bar{c}^\prime\end{array}\right),
\end{equation}
where the symbol ``$\cdot$" denotes the integration over the proper superspace \cite{Tyler}. Moreover, the hessians are given by
\begin{align}
	\label{H_v}
	\mathcal{H}_v&=\frac{1}{4g^2}\left(-\Box f(\Box)+M^2_v\right)\delta^7(z,z^\prime);\\
	\label{H_phi}
	\mathcal{H}_\phi&=\left(\begin{array}{cc}
		\bm{0} & \displaystyle\left(\bm{1}-\frac{\bm{M^{2T}}}{\Box f(\Box)}\right)\bar{D}^2\\
		\displaystyle\left(\bm{1}-\frac{\bm{M^2}}{\Box f(\Box)}\right)D^2 & \bm{0}
	\end{array}\right)\left(\begin{array}{cc}
		\bm{1}\delta_+(z,z^\prime) & \bm{0}\\
		\bm{0} & \bm{1}\delta_-(z,z^\prime)
	\end{array}\right);\\
	\label{H_FP}
	\mathcal{H}_{FP}&=\left(1-\frac{M^2_v}{\Box f(\Box)}\right)\left(\begin{array}{cc}
		\bm{0} & -\sigma_1\bar{D}^2\\
		\sigma_1 D^2 & \bm{0}
	\end{array}\right)\left(\begin{array}{cc}
		\bm{1}\delta_+(z,z^\prime) & \bm{0}\\
		\bm{0} & \bm{1}\delta_-(z,z^\prime)
	\end{array}\right).
\end{align}
Note that the hessians (\ref{H_phi}) and (\ref{H_FP}) were also modified by the HD operator $f(\Box)$. This is a consequence of the gauge choice (\ref{gff}). The hessians above are fundamental for the one- and two-loop computations that we will do next.

\section{One- and two-loop calculations}

By enforcing the constraints (\ref{constr}) on the background superfields, the general structure of the quantum effective action in the $\mathcal{N}=2$, $d=3$ superspace is given by \cite{BK}
\begin{equation}
	\Gamma[\bm{\Phi},\bm{\bar{\Phi}}]=\int d^7zK_{eff}(\bm{\Phi},\bm{\bar{\Phi}})+\left[\int d^5zW_{eff}(\bm{\Phi})+\text{H.c.}\right].
\end{equation}
In this context, the SEP is characterized by two objects: the K\"{a}hler effective potential $K_{eff}$ and chiral effective potential $W_{eff}$. In the K\"{a}hlerian approximation, we focus only on $K_{eff}$. The typical method of calculation of $K_{eff}$ relies on the use of perturbation series in powers of $\hbar$, the so-called loop expansion. Thus, we write
\begin{equation}
\label{4ddefeffpot}
K_{eff}(\bm{\Phi},\bm{\bar{\Phi}})=\sum_{n=0}^\infty\hbar^n K^{(n)}(\bm{\Phi},\bm{\bar{\Phi}}).
\end{equation}
where $K^{(n)}$ denotes the $n$-loop quantum correction. The tree-level EP $K^{(0)}$ can be read from the classical action (\ref{model}) by turning off the gauge superfield $V=0$. Therefore,
\begin{equation}
	\label{zero_loop}
	K^{(0)}(\bm{\Phi},\bm{\bar{\Phi}})=\bm{\bar{\Phi}}\bm{\Phi}.
\end{equation}
Of course, this K\"{a}hler potential for the model (\ref{model}) is identical to the standard one \cite{BK}.  

Now, let us consider the one-loop correction, which can be obtained from (\ref{combined}) by integrating out the quantum superfields. Doing this, we arrive at the expression for the one-loop euclidean effective action \cite{GRU}
\begin{equation}
	\label{1loopEA}
	\Gamma^{(1)}=-\frac{1}{2}\textrm{Tr}\ln\mathcal{H}_v-\frac{1}{2}\textrm{Tr}\ln\mathcal{H}_\phi+\frac{1}{2}\textrm{Tr}\ln\mathcal{H}_{FP}.
\end{equation}
Notice in (\ref{H_v}) that there is no spinor covariant derivative in $\mathcal{H}_v$. Thus, due to properties of the delta function over the Grassmann variables, the first trace in Eq. (\ref{1loopEA}) vanishes. The remaining traces can be handled as follows. Let
\begin{equation}
	\mathcal{C}_\phi=\left(\begin{array}{cc}
		\bm{0} & f(\Box)\bm{1}\bar{D}^2\\
		f(\Box)\bm{1}D^2 & \bm{0}
	\end{array}\right)\left(\begin{array}{cc}
		\bm{1}\delta_+(z,z^\prime) & \bm{0}\\
		\bm{0} & \bm{1}\delta_-(z,z^\prime)
	\end{array}\right)
\end{equation}
be a operator which is independent of the background superfields. Since the contribution of $-\frac{1}{2}\textrm{Tr}\ln(\mathcal{C}_\phi\mathcal{H}_\phi)$ for the SEP differs from  $-\frac{1}{2}\textrm{Tr}\ln\mathcal{H}_\phi$ by an additive constant, this allows us to redefine $\mathcal{H}_\phi$ such that $\mathcal{H}_\phi\rightarrow\mathcal{C}_\phi\mathcal{H}_\phi$, where $\mathcal{C}_\phi\mathcal{H}_\phi$ is a block diagonal matrix. Thus,
\begin{align}
	\label{matter_cont}
	-\frac{1}{2}\textrm{Tr}\ln(\mathcal{C}_\phi\mathcal{H}_\phi)&=-\frac{1}{2}\textrm{Tr}\ln\left(\begin{array}{cc}
		\Box f(\Box)\bm{1}-\bm{M^2} & \bm{0}\\
		\bm{0} & \Box f(\Box)\bm{1}-\bm{M^{2T}}
	\end{array}\right)\nonumber\\
	&=-\frac{1}{2}\textrm{Tr}_+\ln\left(\Box f(\Box)\bm{1}-\bm{M^2}\right)-\frac{1}{2}\textrm{Tr}_-\ln\left(\Box f(\Box)\bm{1}-\bm{M^{2T}}\right)\nonumber\\
	&=-\frac{1}{2}\textrm{Tr}_+\ln\left(\Box f(\Box)\bm-M^2_v\right)+\text{H.c.} \ ,
\end{align}
where we have used the eigenvalues of the matrix $\bm{M^2}$ (and $\bm{M^{2T}}$), which are $\lambda_1=0$ and $\lambda_2=M^2_v$. Moreover, $\text{Tr}_+$ and $\text{Tr}_-$ denote the traces over the chiral and antichiral subspaces, respectively.

Similarly, we can introduce the operator
\begin{equation}
	\mathcal{C}_{FP}=\left(\begin{array}{cc}
		\bm{0} & f(\Box)\sigma_1\bar{D}^2\\
		-f(\Box)\sigma_1D^2 & \bm{0}
	\end{array}\right)\left(\begin{array}{cc}
		\bm{1}\delta_+(z,z^\prime) & \bm{0}\\
		\bm{0} & \bm{1}\delta_-(z,z^\prime)
	\end{array}\right),
\end{equation}
so that the ghost contribution corresponds to
\begin{align}
	\label{ghost_cont}
	\frac{1}{2}\textrm{Tr}\ln(\mathcal{C}_{FP}\mathcal{H}_{FP})&=\frac{1}{2}\textrm{Tr}\ln\left(\begin{array}{cc}
		\left(\Box f(\Box)\bm-M^2_v\right)\bm{1} & \bm{0}\\
		\bm{0} & \left(\Box f(\Box)\bm-M^2_v\right)\bm{1}
	\end{array}\right)\nonumber\\
	&=\textrm{Tr}_+\ln\left(\Box f(\Box)\bm-M^2_v\right)+\text{H.c.} \ .
\end{align}
Therefore, substituting (\ref{matter_cont}) and (\ref{ghost_cont}) into (\ref{1loopEA}), we find
\begin{equation}
	\label{HD_trace}
	\Gamma^{(1)}=\frac{1}{2}\textrm{Tr}_+\ln\left(\Box f(\Box)\bm-M^2_v\right)+\text{H.c.} \ .
\end{equation}
Now we did all the $D$-algebra, the next step is to factor the HD operator $\Box f(\Box)-M^2_v$ into the product of standard wave operators $\Box-m^2_k$. This aim can be achieved by invoking the fundamental theorem of algebra. If $N$ denotes the degree of the polynomial $P(z)=zf(z)-M^2_v$, then we can write
\begin{equation}
	\label{poly}
	P(z)=\sum_{k=0}^N a_kz^k=a_N\prod_{k=1}^N\left(z-m^2_k\right),
\end{equation}
where $a_N$ is a constant and $m^2_k$ are the zeroes of $P(z)$, which are background-dependent. Additionally, we have assumed that all $m_k$ are distinct, real and positive for the sake of simplicity. This assumption can be accomplished through an appropriate choice of the coefficients of $P(z)$.

In view of the discussion above, let us split the trace (\ref{HD_trace}) into $N$ traces involving wave operators $\Box-m^2_k$:
\begin{equation}
	\Gamma^{(1)}=\frac{1}{2}\sum_{k=1}^N\textrm{Tr}_+\ln\left(\Box-m^2_k\right)+\text{H.c.} \ .
\end{equation}
Finally, in order to evaluate these traces, we can follow the same approach as in \cite{BMS}. Therefore, it is possible to show that the one-loop SEP is
\begin{equation}
	\label{one-loop}
	K^{(1)}(\bm{\Phi},\bm{\bar{\Phi}})=-\frac{1}{2\pi}\sum_{k=1}^N m_k .
\end{equation}
Thus, we find that the one-loop SEP for the HD model (\ref{model}) is finite. Indeed, one-loop ultraviolet finiteness is a typical characteristic of three-dimensional gauge theories \cite{3d}.

\begin{figure}[!h]
	\begin{center}
		\includegraphics[angle=0,scale=0.4]{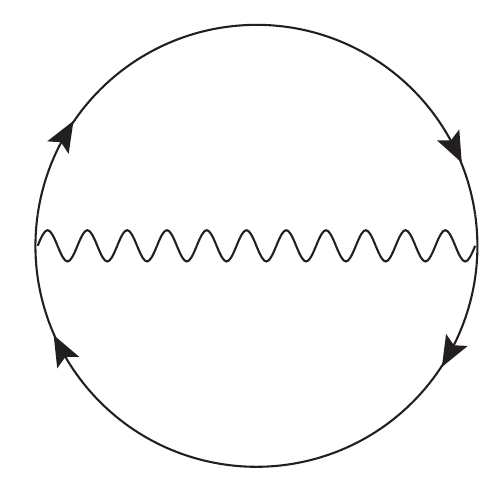}
	\end{center}
	\caption{The only non-zero contribution to the two-loop SEP.}
	\label{fig:non-zero}
\end{figure}

Let us now move on to the calculation of the two-loop SEP. To do this we need to determine the propagators by inverting the hessians (\ref{H_v}-\ref{H_FP}). Therefore, we obtain
\begin{align}
	\label{prop_v}
	G_v(z,z^\prime)&=A(\Box)\delta^7(z,z^\prime);\\
	\label{prop_phi}
	G_\phi(z,z^\prime)&=\left(\begin{array}{cc}
		\bm{0} & \left(B(\Box)\bm{P_1}-\Box^{-1}\bm{P_2}\right)\bar{D}^2\delta_-(z,z^\prime)\\
		\left(B(\Box)\bm{P_1^T}-\Box^{-1}\bm{P_2^T}\right)D^2\delta_+(z,z^\prime) & \bm{0}
	\end{array}\right);\\
\label{prop_FP}
	G_{FP}(z,z^\prime)&=B(\Box)\left(\begin{array}{cc}
		\bm{0} & -\sigma_1\bar{D}^2\delta_-(z,z^\prime)\\
		\sigma_1 D^2\delta_+(z,z^\prime) & \bm{0}
	\end{array}\right),
\end{align}
where we have used the following matrix projection operators \cite{NN}
\begin{equation}
	\label{projection}
	\bm{P_1}=\frac{1}{\bm{\bar{\Phi}}\bm{\Phi}}\left(\begin{array}{c}
		\Phi_+\\
		-\Phi_-
	\end{array}\right)\left(\begin{array}{cc}
		\bar{\Phi}_+ & -\bar{\Phi}_-
	\end{array}\right); \ \ \ \bm{P_2}=\frac{1}{\bm{\bar{\Phi}}\bm{\Phi}}\left(\begin{array}{c}
		\bar{\Phi}_-\\
		\bar{\Phi}_+
	\end{array}\right)\left(\begin{array}{cc}
		\Phi_- & \Phi_+
	\end{array}\right).
\end{equation}
Additionally,
\begin{equation}
	\label{AB}
	A(\Box)=\frac{4g^2}{\Box f(\Box)-M_v^2}; \ \ \ B(\Box)=-\frac{f(\Box)}{\Box f(\Box)-M_v^2}.
\end{equation}
It is worth noticing that despite the HD operator $f(\Box)$ appearing in all propagators above, only (\ref{prop_v}) is improved at high momentum due to $f(\Box)$, the other ones (\ref{prop_phi}) and (\ref{prop_FP}) go as $-\Box^{-1}$ at high momentum. This implies that supergraphs with vertices connected by only matter and ghost propagators should be divergent. However, such supergraphs are not allowed by the propagators (\ref{prop_phi}-\ref{prop_FP}) and cubic vertices originating from the functional (\ref{ghost_3}). On the other hand, there are five supergraphs which are allowed by the propagators (\ref{prop_v}-\ref{prop_phi}) and vertices originating from (\ref{cubic}-\ref{quartic}). From all these supergraphs, only the one drawn in Fig. \ref{fig:non-zero} is not zero, because after the $D$-algebra we are left with exactly two $D$'s and two $\bar{D}$'s acting on one of the $\delta$-functions.

The expression for the two-loop correction to the euclidean effective action corresponding to the supergraph shown in Fig. \ref{fig:non-zero} is given by
\begin{align}
	\label{expression}
	\Gamma^{(2)}&=\int d^7z_1 d^5\bar{z}_2d^5z_3d^7z_4d^5\bar{z}_5d^5z_6\frac{\delta^3 S^{(3)}}{\delta v(z_1)\delta\bar{\phi}_i(z_2)\delta\phi_j(z_3)}\frac{\delta^3 S^{(3)}}{\delta v(z_4)\delta\bar{\phi}_k(z_5)\delta\phi_l(z_6)}\nonumber\\
	&\times G_v(z_1,z_4)(G_{-+})_{il}(z_2,z_6)(G_{+-})_{jk}(z_3,z_5),
\end{align}
where $G_{-+}$ and $G_{+-}$ are matrix propagators lying along the anti-diagonal of $G_\phi$ [see Eq. (\ref{prop_phi})], and $i,j,k,l$ are matrix indices. A straightforward computation using (\ref{cubic}) lead us to the following three-point vertex
\begin{equation}
	\label{calc_vert}
	\frac{\delta^3 S^{(3)}}{\delta v(z_1)\delta\bm{\bar{\phi}}_i(z_2)\delta\bm{\phi}_j(z_3)}=(\sigma_3)_{ij}\delta_-(z_1,z_2)\delta_+(z_1,z_3).
\end{equation}
Substituting (\ref{prop_v}) and (\ref{calc_vert}) into (\ref{expression}), and then using the anti- and chiral delta functions to evaluate the integrals, we find 
\begin{equation}
	\label{expression_2}
	\Gamma^{(2)}=\int d^7z_1d^7z_4(\sigma_3)_{ij}(\sigma_3)_{kl}\left[A(\Box_1)\delta^3(x_1,x_4)\right]\delta^4(\theta_1,\theta_4)(G_{-+})_{il}(z_1,z_4)(G_{+-})_{jk}(z_1,z_4).
\end{equation}
By means of the identities
\begin{align}
	\delta^4(\theta_1,\theta_4)D_1^2\delta_+(z_1,z_4)=\delta^7(z_1,z_4); \ \ \ \delta^4(\theta_1,\theta_4)\bar{D}_1^2\delta_-(z_1,z_4)=\delta^7(z_1,z_4),
\end{align}
we are able to easily prove that
\begin{align}
	\delta^4(\theta_1,\theta_4)(G_{-+})_{il}(z_1,z_4)&=\left[B(\Box_1)\left(\bm{P_1^T}\right)_{il}-\Box_1^{-1}\left(\bm{P_2^T}\right)_{il}\right]\delta^3(x_1,x_4)\delta^4(\theta_1,\theta_4);\\
	\delta^4(\theta_1,\theta_4)(G_{+-})_{jk}(z_1,z_4)&=\left[B(\Box_1)\left(\bm{P_1}\right)_{jk}-\Box_1^{-1}\left(\bm{P_2}\right)_{jk}\right]\delta^3(x_1,x_4)\delta^4(\theta_1,\theta_4).
\end{align}
Therefore, after a little algebraic manipulation, the expression (\ref{expression_2}) can be rewritten as
\begin{align}
	\label{expression_3}
	\Gamma^{(2)}&=\int d^4\theta_1d^3x_1d^3x_4\Big\{\text{Tr}\left(\bm{P_1}\sigma_3\bm{P_1}\sigma_3\right)\left[A(\Box_1)\delta^3(x_1,x_4)\right]\left[B(\Box_1)\delta^3(x_1,x_4)\right]^2\nonumber\\
	&+2\text{Tr}\left(\bm{P_1}\sigma_3\bm{P_2}\sigma_3\right)\left[A(\Box_1)\delta^3(x_1,x_4)\right]\left[B(\Box_1)\delta^3(x_1,x_4)\right]\left[-\Box_1^{-1}\delta^3(x_1,x_4)\right]\nonumber\\
	&+\text{Tr}\left(\bm{P_2}\sigma_3\bm{P_2}\sigma_3\right)\left[A(\Box_1)\delta^3(x_1,x_4)\right]\left[-\Box_1^{-1}\delta^3(x_1,x_4)\right]^2\Big\}.
\end{align}
The traces above are evaluated using the definitions (\ref{projection}), and the results can be expressed in terms of the matrix notation (\ref{matrix_not}). Thus,
\begin{align}
	\label{traces}
	\text{Tr}\left(\bm{P_1}\sigma_3\bm{P_1}\sigma_3\right)=\text{Tr}\left(\bm{P_2}\sigma_3\bm{P_2}\sigma_3\right)=\left(\frac{\bm{\bar{\Phi}}\sigma_3\bm{\Phi}}{\bm{\bar{\Phi}}\bm{\Phi}}\right)^2; \ \ \ \text{Tr}\left(\bm{P_1}\sigma_3\bm{P_2}\sigma_3\right)=\left|\frac{\bm{\Phi^T}\sigma_1\bm{\Phi}}{\bm{\bar{\Phi}}\bm{\Phi}}\right|^2.
\end{align}
In order to get Feynman integrals with known solutions in terms of elementary functions, we need once again to factor the HD operator $\Box f(\Box)-M^2_v$ into the product of standard wave operators $\Box-m^2_k$ using the fundamental theorem of algebra (\ref{poly}). This allows us to find for $A(\Box)$ and $B(\Box)$, defined in Eq. (\ref{AB}), the following partial fraction representations  \cite{MH}
\begin{align}
	\label{partial_dec_1}
	A(\Box)&=\frac{4g^2a_N^{-1}}{\prod_{k=1}^N\left(\Box-m_k^2\right)}=4g^2a^{-1}_N\sum_{k=1}^N\frac{c_k}{\Box-m_k^2};\\
	\label{partial_dec_2}
	B(\Box)&=-\frac{a_N^{-1}f(\Box)}{\prod_{k=1}^N\left(\Box-m_k^2\right)}=-a^{-1}_N\sum_{k=1}^N\frac{d_k}{\Box-m_k^2},
\end{align}
where the residues $c_k$ and $d_k$ are given by \cite{Lang}
\begin{equation}
	\label{residues}
	c_k=\prod_{j\neq k}\frac{1}{m_k^2-m_j^2}; \ \ \ d_k=\prod_{j\neq k}\frac{f(m_k^2)}{m_k^2-m_j^2}.
\end{equation}
Inserting Eqs. (\ref{traces}-\ref{partial_dec_2}) into (\ref{expression_3}), and then passing to the momentum space, we arrive at
\begin{align}
	\label{expression_4}
	\Gamma^{(2)}&=-4g^2a_N^{-1}\int d^7z\int\frac{d^3pd^3q}{(2\pi)^6}\Bigg\{\left(\frac{\bm{\bar{\Phi}}\sigma_3\bm{\Phi}}{\bm{\bar{\Phi}}\bm{\Phi}}\right)^2\bigg[a_N^{-2}\sum_{k=1}^N\sum_{l=1}^N\sum_{m=1}^N c_k d_l d_m\frac{1}{p^2+m_k^2}\frac{1}{q^2+m_l^2}\nonumber\\
	&\times\frac{1}{(p+q)^2+m_m^2}+\sum_{k=1}^N c_k\frac{1}{p^2+m_k^2}\frac{1}{q^2}\frac{1}{(p+q)^2}\bigg]+2\left|\frac{\bm{\Phi^T}\sigma_1\bm{\Phi}}{\bm{\bar{\Phi}}\bm{\Phi}}\right|^2a_N^{-1}\sum_{k=1}^N\sum_{l=1}^N c_k d_l\nonumber\\&\times\frac{1}{p^2+m_k^2}\frac{1}{q^2+m_l^2}\frac{1}{(p+q)^2}\Bigg\}.
\end{align}
Even though perturbative quantum contributions to the effective action are in general highly non-local in the coordinates $x$ \cite{Rivers}, notice that we have ended up with a contribution (\ref{expression_4}) that is local. This is due to the fact that the derivatives of the background superfields with respect to $x$ have been neglected in our computations [see Eq. (\ref{constr})].

The 2-loop vacuum integrals with arbitrary masses which appear in Eq. (\ref{expression_4}) also arise in the $3d$ physics of the electroweak phase transition \cite{FKRS} and spontaneous breaking of the gauge symmetry in the Maxwell-Chern-Simons theory \cite{TTH}. They can be solved by means of the formula
\begin{equation}
	\label{integral}
	\int \frac{d^3pd^3q}{(2\pi)^6}\frac{1}{p^2+m^2_1}\frac{1}{q^2+m^2_2}\frac{1}{(p+q)^2+m^2_3}=\frac{1}{32\pi^2}\left[\frac{1}{2\varepsilon}+1-2\ln\left(\frac{m_1+m_2+m_3}{\bar{\mu}}\right)\right],
\end{equation}
where $\varepsilon=\frac{1}{2}(3-d)$ and $\bar{\mu}$ is an arbitrary mass parameter.

It is worth pointing out that despite the singularity $\varepsilon^{-1}$ arising in all integrals above, they are not of concern because they cancel in the sums due to the identity
\begin{equation}
	\label{iden}
	\sum_{k=1}^Nc_k=0.
\end{equation}
Finally, substituting (\ref{integral}) into (\ref{expression_4}), and then using the identity (\ref{iden}), we can infer that the two-loop SEP is
\begin{align}
	\label{two-loop}
	K^{(2)}(\bm{\Phi},\bm{\bar{\Phi}})&=\frac{g^2a_N^{-1}}{4\pi^2}\Bigg\{\left(\frac{\bm{\bar{\Phi}}\sigma_3\bm{\Phi}}{\bm{\bar{\Phi}}\bm{\Phi}}\right)^2\bigg[a_N^{-2}\sum_{k=1}^N\sum_{l=1}^N\sum_{m=1}^N c_k d_l d_m\ln\left(\frac{m_k+m_l+m_m}{\bar{\mu}}\right)\nonumber\\
	&+\sum_{k=1}^N c_k\ln\left(\frac{m_k}{\bar{\mu}}\right)\bigg]+2\left|\frac{\bm{\Phi^T}\sigma_1\bm{\Phi}}{\bm{\bar{\Phi}}\bm{\Phi}}\right|^2a_N^{-1}\sum_{k=1}^N\sum_{l=1}^N c_k d_l\ln\left(\frac{m_k+m_l}{\bar{\mu}}\right)\Bigg\},
\end{align}
where the residues $c_k$, $d_k$, and the masses $m_k$ are functions of the background superfields. Contrary to the one-loop polynomial SEP (\ref{one-loop}), the two-loop one (\ref{two-loop}) has a logarithmic behavior in the masses. Moreover, in contrast to the two-loop divergent SEP for standard three-dimensional gauge theories \cite{3d}, the one for the HD model (\ref{model}) is finite. 

As an illustration of our general results, let us consider a HD model (\ref{model}) whose explicit form of the HD operator $f(\Box)$ is defined as
\begin{equation}
	\label{example}
	f(\Box)=1-\frac{\Box}{\Lambda^2},
\end{equation}
where $\Lambda>0$. In particular, when we set the parameter to be infinitely large $\Lambda\to\infty$, we recover the usual $\mathcal{N}=2$, $d=3$ SQED.

It follows from the example (\ref{example}) that we have to find all zeroes of the polynomial [see Eq. (\ref{poly})]
\begin{equation}
	\label{equation}
	-\frac{z^2}{\Lambda^2}+z-4g^2\bm{\bar{\Phi}}\bm{\Phi}=0.
\end{equation}
The degree $N$ of this polynomial and the coefficient $a_2$ are given by
\begin{equation}
	\label{deg_coe}
	N=2; \ \ \ a_2=-\frac{1}{\Lambda^{2}}.
\end{equation}
If $\Lambda^2>16g^2\bm{\bar{\Phi}}\bm{\Phi}$, then the quadratic equation (\ref{equation}) admits two distinct real solutions:
\begin{equation}
	\label{masses}
	m_{\pm}^2=\frac{\Lambda^2}{2}\left(1\pm\sqrt{1-\frac{16g^2\bm{\bar{\Phi}}\bm{\Phi}}{\Lambda^2}}\right),
\end{equation}
where $m_{+}^2$ is the mass of the Ostrogradsky ghost (which is inevitable in any local HD theory) and $m_{-}^2$ is the mass acquired by the gauge superfield. Indeed, notice that $m_{-}^2=0$ for a vanishing background superfield, $\bm{\Phi}=0$, and $m_{-}^2\simeq M^2_v$ for a large mass scale, $\Lambda\to\infty$.

The square roots are
\begin{equation}
	\label{roots}
	m_{\pm}=\frac{\Lambda}{\sqrt{2}}\sqrt{1\pm\sqrt{1-\frac{16g^2\bm{\bar{\Phi}}\bm{\Phi}}{\Lambda^2}}}.
\end{equation}
To find the one-loop SEP, one can just insert Eqs. (\ref{deg_coe}) and (\ref{roots}) into (\ref{one-loop}). On the other hand, to obtain the two-loop SEP, it is still necessary to determine the residues (\ref{residues}). Substituting (\ref{roots}) into (\ref{residues}) and using the definition (\ref{example}), we get
\begin{equation}
	\label{res}
	c_\pm=\pm\frac{1}{\Lambda\sqrt{\Lambda^2-16g^2\bm{\bar{\Phi}}\bm{\Phi}}}; \ \ \ d_{\pm}=\pm\frac{\Lambda\mp\sqrt{\Lambda^2-16g^2\bm{\bar{\Phi}}\bm{\Phi}}}{2\Lambda^2\sqrt{\Lambda^2-16g^2\bm{\bar{\Phi}}\bm{\Phi}}}.
\end{equation}
Finally, one can just insert Eqs. (\ref{deg_coe}), (\ref{roots}), and (\ref{res}) into (\ref{two-loop}) to find the two-loop SEP.

Even though the formula (\ref{two-loop}) is valid for $f(\Box)$ with arbitrary degree, it becomes very unwieldy to write out in full when $f(\Box)$ is a polynomial with degree higher than two.

\section{Conclusions}

In this work, we formulated a higher-derivative generalization of the $\mathcal{N}=2$, $d=3$ supersymmetric quantum electrodynamics, with the introduction of a higher-derivative operator in the gauge sector, where such operator is a polynomial function of the d'Alembertian with arbitrary degree. For this theory, we found explicit expressions for the one- and two-loop superfield effective potentials. All quantum calculations were performed using the background field approach in a higher-derivative $R_\xi$ gauge to remove the unwanted mixed terms between the quantum superfields.

The two main results of our paper are as follows. First, the derivation of an explicit exact expression for the one-loop superfield effective potential (\ref{one-loop}). This result is relevant because it improves the result we found in our earlier work \cite{GGNPS}, where the one-loop superfield effective potential was given in terms of integrals over the momenta. Second, the full derivation of the two-loop superfield effective potential (\ref{two-loop}) in a closed form and in terms of elementary functions. The importance of this result lies in the fact that, in contrast our earlier study \cite{GNP2}, it is the first time that these two-loop corrections were determined for a gauge superfield theory with higher derivatives. It is worth mentioning that the functional structures of (\ref{one-loop}) and (\ref{two-loop}) are highly dependent of the masses and residues of the propagators (\ref{partial_dec_1}) and (\ref{partial_dec_2}) that, in addition to degrees of freedom associated with the massive vector multiplet, describe extra degrees of freedom associated with Ostrogradsky ghosts. This non-trivial dependence of the SEP on the masses and residues suggests that the ground state of the theory can be modified by the presence of the ghosts. Therefore, our results (\ref{one-loop}) and (\ref{two-loop}) indicate that the Ostrogradsky ghosts can affect the ground state of the theory and the physical phenomena related to it, such as the spontaneous symmetry breaking.

The most natural continuation of this work would consist in extending the results obtained here for a four-dimensional version of the higher-derivative supersymmetric quantum electrodynamics (\ref{model}). In this case, if we insist on working with the model defined in (\ref{model}), the two-loop SEP will not be finite. This occurs due to the fact that the two-loop vacuum integral in four dimensions leads to involved divergences [see Eq. (4.20) of Ref. \cite{FJJ}] and these divergences are not completely canceled due to the identity (\ref{iden}), in contrast to the divergences generated by the three-dimensional integral (\ref{integral}). Therefore, it is obligatory to include higher-derivative operators into the matter sector of (\ref{model}) in order to ensure the finiteness of the two-loop SEP in four dimensions. However, the calculations become technically more challenging due to the more involved structure of the propagators and the large number of vertices. For this reason, we expect to carry out these studies in a future work.

\vspace{5mm}

{\bf Acknowledgments.} The author would like to express his gratitude to A. Freitas for valuable discussions.

\end{document}